\pdfoutput=1
\RequirePackage{fixltx2e}

\documentclass[reprint,aip,pop,amsmath,amsfonts,amssymb,floatfix]{revtex4-1}

\usepackage{graphicx,xcolor,latexsym}

\usepackage[colorlinks=true, linkcolor=blue, citecolor=blue, urlcolor=blue, bookmarks=true, breaklinks=true,
            pdftitle={Molecular Dynamics Simulations of Field Emission From a Planar Nanodiode}, 
            pdfauthor={Kristinn Torfason}]{hyperref}
\usepackage[all]{hypcap}
\usepackage{bookmark}

\newcommand{\ud}{\mathrm{d}}

\begin{document}


\title{Molecular Dynamics Simulations of Field Emission From a Planar Nanodiode}


\author{Kristinn Torfason}

\author{Agust Valfells}

\author{Andrei Manolescu}
\affiliation{School of Science and Engineering, Reykjavik University, Menntavegur 1, IS-101 Reykjavik, Iceland}

\date{\today}

\begin{abstract}
High resolution molecular dynamics simulations with full Coulomb interactions of electrons
are used to investigate field emission in planar nanodiodes. The effects
of space-charge and emitter radius are examined and
compared to previous results concerning transition from Fowler-Nordheim
to Child-Langmuir current~\cite{Lau1870603, Feng2006}.  The
Fowler-Nordheim law is used to determine the current density injected into
the system and the Metropolis-Hastings algorithm to find a favourable
point of emission on the emitter surface. A simple fluid like model is also developed
and its results are in qualitative agreement with the simulations.
\end{abstract}


\maketitle


\section{Introduction}
Field emission of electrons from a metallic surface~\cite{Fowler01051928, Forbes08112007, Jensen1528} is an important process in vacuum electronics~\cite{eichmeier2008vacuum, zhu2004vacuum}.
Its primary use is for cold cathodes in applications such as microwave tubes, electron microscopes and flat panel displays.
In addition to this beneficial application of field emission there are also negative implications,
such as vacuum breakdown at high field strengths and dielectric surface breakdown initiated by field emitted electrons~\cite{neuber2001windows}.

The physical basis for field emission is that a strong applied electric field can deform the surface barrier of a
metal-vacuum interface so as to increase the tunnelling probability to such a degree that a considerable current of electrons can be drawn from the surface.
This process was first described by Fowler and Nordheim~\cite{Fowler01051928} and has since been recast and extended for different applications.
A particular area of interest is the effect of space-charge on field emission.  Barbour et~al. addressed this issue in 1953~\cite{PhysRev.92.45},
while Lau et~al. conducted a study of transition from field emission to space-charge dominated emission in 1994~\cite{Lau1870603}.
In both cases the approach was based on finding the equilibrium injection current that corresponds to the space-charge modified surface field.
Feng and Verboncoeur~\cite{Feng2006} used a particle-in-cell code to simulate this problem,
in which they used a detailed model that takes into account the effect of image-charge on the potential barrier for field emission.
Rokhlenko et~al.~\cite{Rokhlenko3272690} developed an elegant approach to calculate the effect of space-charge on field emission in a one dimensional system
that included lowering of the potential barrier by an effective work function, that matched the results of Feng and Verboncoeur quite well.
Other recent work includes a three dimensional theory for space-charge effects on field emission from a Spindt type emitter~\cite{Jensen14905}
and on space-charge and quantum effects~\cite{Jensen3692577}.

The work presented in this paper was motivated by the desire to use molecular dynamics methods to simulate electron beams in vacuum micro- and nanoelectronics devices.
An advantage of this approach is that it can account more accurately for collisional effects than simulations based on fluid models or particle-in-cell codes.
A disadvantage is the high computational cost, but in the small systems that are of interest, the number of free electrons is small enough that this is not an issue.
Previous work in this area~\cite{PhysRevLett.104.175002, Jonsson4793451, 6979259} has been based on something akin to photoemission as the source of electrons,
whereby the surface electric field does not directly affect the rate of electron emission from the cathode although space-charge effects do limit the current via virtual cathode formation.
Due to the importance of field emission it is desirable to implement that type of emission process in the molecular dynamics code being used for
simulation of vacuum nanoelectronics devices.

In this paper we present a field emission model based on image-charge considerations that is suitable for the molecular dynamics approach.
We show how this model can replicate already established results, particularly with regard to space-charge effects, and we use our model to simulate field emission in a planar diode of limited emitter area.
The simulation results are then compared to a simple fluid model for field emission from a planar emitter of finite area.

\autoref{sec:method} of this paper gives a description of the model and simulation methodology used.
Simulation results are presented in~\autoref{sec:results} followed by a fluid model description of two-dimensional effects in field emission
in~\autoref{sec:fluid} and finally by a short summary and discussion in~\autoref{sec:summary}.

\section{Methodology\label{sec:method}}
\begin{figure}[bt]
    \centering
    \includegraphics{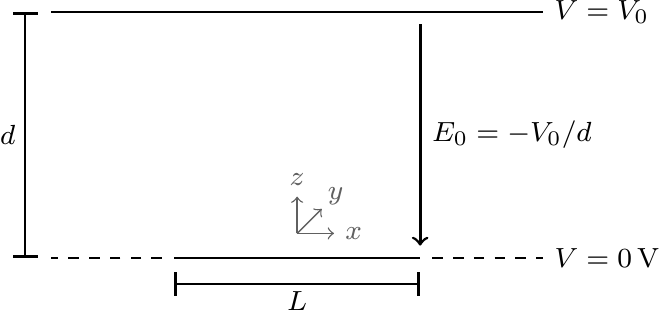}
    \caption{The model of the planar nanodiode.}
    \label{fig:system}
  \end{figure}

  The model used is a planar vacuum diode with gap width
  \(d\) as is depicted in~\autoref{fig:system}.  The potential at the
  cathode is zero and the anode potential is \(V_0\).  Field emission
  is only allowed to take place from a finite square-shaped area on the
  cathode surface. Both the cathode and anode surfaces are infinite.
  The side length, \(L\), of this square emitter
  region is smaller than the gap spacing \(d\).  A molecular dynamics (MD)
  approach is used to calculate electron motion and is the basis for
  the field emission algorithm.  The simulation is of high resolution
  in the sense that every single electron present in the vacuum gap
  is treated as an individual particle.  The Coulomb field due to
  every electron in the system along with the image-charge partners on
  either side of the cathode/anode are taken into account.  Thus the
  total field at any point is \(E = E_{sc} + E_0\), where \(E_{sc}\)
  is the detailed space-charge field and \(E_0\) is the vacuum field.
  Particle advancement is calculated using Verlet integration with a
  time step of \(0.1\,\mathrm{fs}\).

\begin{figure}[tb]
    \centering
    \includegraphics{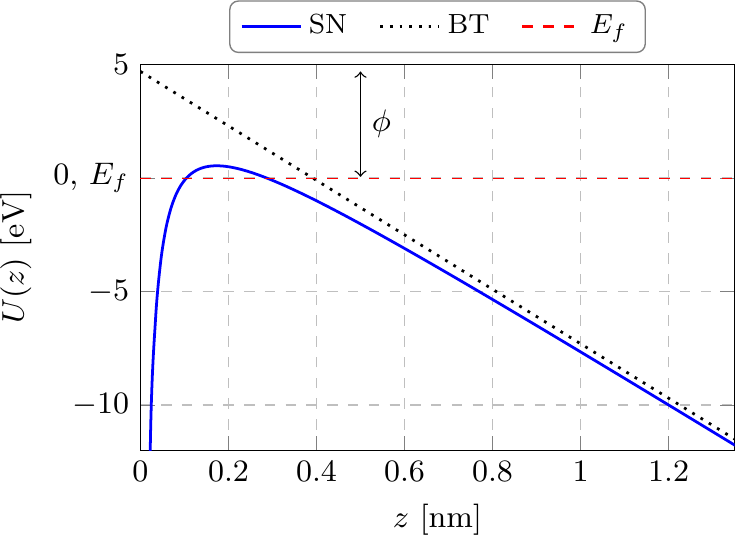}
    \caption{The bare triangular barrier vs. the Schottky-Nordheim barrier for field emission.}
    \label{fig:barrier}
  \end{figure}

  Field emission is a quantum mechanical tunnelling process which can be described with the Fowler-Nordheim equation~\cite{Fowler01051928}
  \begin{equation}\label{eq:FN-eq}
    J = \frac{A}{t^2(\ell)\phi}F^2 \mathrm{e}^{-\nu(\ell)B\phi^{\frac{3}{2}}/F}\, ,
  \end{equation}
  where \(\phi\) is the work-function and \(F\) is the field
  at the surface of the cathode, taken to be positive. \(A = e^2/(16\pi^2\hbar)\;
  [\mathrm{A}\,\mathrm{eV}\,\mathrm{V}^{-2}]\)
  and \(B = 4/(3\hbar) \sqrt{2m_e e}\;
  [\mathrm{eV}^{-\frac{3}{2}}\,\mathrm{V}\,\mathrm{m}^{-1}]\) are the
  first and second Fowler-Nordheim constants, while \(\nu(\ell)\)
  is called the Nordheim function and arises due to the
  image-charge effect.  It contains complete elliptic integrals of
  the first and second kind and is related to \(t(\ell)\) by the
  relation \(t(\ell) = \nu(\ell) - (4/3)\ell\, \ud \nu(\ell) / \ud
  \ell\). Approximations found by Forbes and Deane~\cite{Forbes08112007},
  \begin{subequations}\label{eq:nordheim-fun}
  \begin{equation}
  \nu(\ell) = 1 - \ell + \frac{1}{6} \ell \ln(\ell)
  \end{equation}
  and
  \begin{equation}
  t(\ell) = 1 + \ell\left(\frac{1}{9} - \frac{1}{18}\ln(\ell) \right)
  \end{equation}
  are used, where
  \begin{equation}\label{eq:ell}
    \ell = \frac{e}{4\pi\varepsilon_0}\frac{F}{\phi^2}\,.
  \end{equation}
  \end{subequations}
  Image-charge is taken into consideration with two objectives in mind. First,
  to maintain the proper boundary conditions at the cathode we include
  image-charge partners for the space-charge to be found in the gap. Second,
  the image-charge of the electron being emitted is taken into account in
  calculating the barrier potential which the electron must tunnel through.
  This can be seen in~\autoref{fig:barrier} where the bare triangular (BT)
  barrier is given by \(U^{BT}(z) = \phi - eFz\)
  and the screened Schottky-Nordheim (SN) barrier by
  \(U^{SN}(z) = \phi - eFz - e^2/(16\pi\epsilon_0 z)\).
  The SN barrier contains an additional term which arises due to the image-charge effect.
  Tunnelling is more pronounced due to this term because the barrier height and width are reduced compared to the triangular barrier.

  Numerical methods are used to obtain the dynamics in the system. The total number of electrons in the system is not constant since electrons are continuously
  entering the system at the cathode and leaving at the anode.
  The Metropolis-Hastings algorithm~\cite{Hastings} is used to sample the surface electric field to find a favourable locations for emission.
  The probability of emission \(D_F = \exp(-\nu(\ell)B\phi^{\frac{3}{2}}/F)\), is then evaluated for small areas at those locations in order to determine
  whether emission occurs.
  The current density emitted from the cathode is normalized such that Fowler-Nordheim current density is obtained.
  Relativistic effects and radiation caused by acceleration of electrons are safely neglected as all occurring velocities are much smaller than the speed of light.
\section{Results of the MD simulations\label{sec:results}}
  \subsection{Current-voltage characteristic}
\begin{figure}
    \centering
    \includegraphics{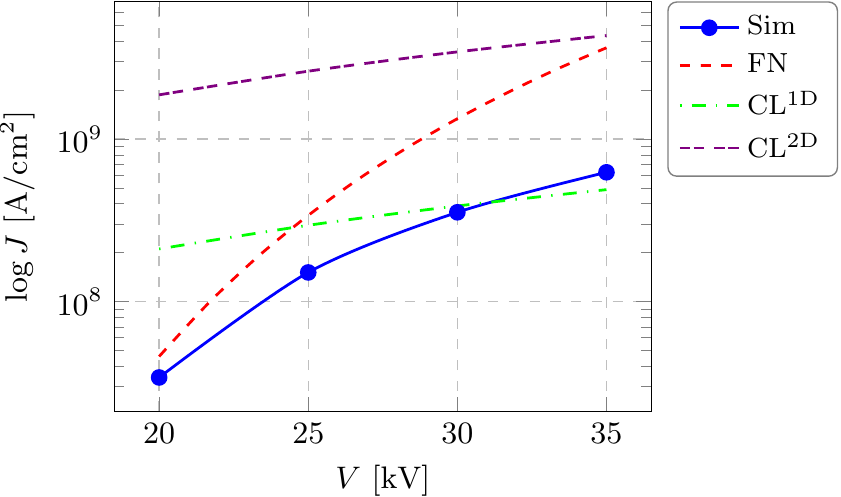}
    \caption{The current density plotted on a log scale vs. the voltage in the system.
    The \textbf{\textcolor{red}{red}} dashed curve shows the results using the value of the vacuum field in the Fowler-Nordheim equation, and
    the \textbf{\textcolor{blue}{blue}} solid curve shows the results from the simulations.
    While the \textbf{\textcolor{green}{green}} dot dashed and \textbf{\textcolor{violet}{violet}} double dashed curves represent the 1D and 2D~\cite{PhysRevLett.87.278301}
    CL limits respectively. The parameters used in the simulation were
    \(\Phi = 4.7\,\mathrm{eV}\), \(d = 2500\,\mathrm{nm}\) and \(L = 100\,\mathrm{nm}\).}
    \label{fig:JvsV}
  \end{figure}

  We start be examining a system with a work function of \(\phi = 4.7\,\mathrm{eV}\). The value is chosen because many metals have a work function in the range
  from \(4.5\,\mathrm{eV}\) to \(5.0\,\mathrm{eV}\). This work function might represent Copper~(Cu), Tungsten~(W) or other metals.
  The gap spacing \(d = 2500\,\mathrm{nm}\) and voltage \(V = 20\text{--}35\,\mathrm{kV}\) were then selected such that the vacuum field was sufficient to obtain field emission.
  Care was taken in selecting the vacuum field such that the parameter \(\ell\) in~\autoref{eq:ell} was less than \(1\). If \(\ell\) is larger than \(1\) then the barrier
  in~\autoref{fig:barrier} will be below the Fermi energy.
  A higher work function means that a higher surface field is required to obtain field emission.
  In~\autoref{fig:JvsV} the current density is plotted as a function of the voltage on a log scale. It shows the exponential increase as is expected from the Fowler-Nordheim
  equation. The \textcolor{red}{red} dashed line shows the Fowler-Nordheim theory (\autoref{eq:FN-eq}) without space-charge effects.
  The \textcolor{blue}{blue} solid line represents the results from the simulations in the steady state. We see that it is lower than the values given by the Fowler-Nordheim theory.
  This is due to space-charge effects lowering the surface field at the cathode and reducing the emission.
  The \textcolor{green}{green} dot dashed curve shows the 1D Child-Langmuir (CL)~\cite{PhysRevSeriesI.32.492,PhysRev.2.450} limit and the double dashed \textcolor{violet}{violet} is the 2D CL limit derived
  by Lau~\cite{PhysRevLett.87.278301}. We see that under high voltage it is possible to go over the 1D CL limit but the 2D limit is still far off.

  \subsection{Surface field at the cathode}

\begin{figure}[t]
    \centering
    \includegraphics{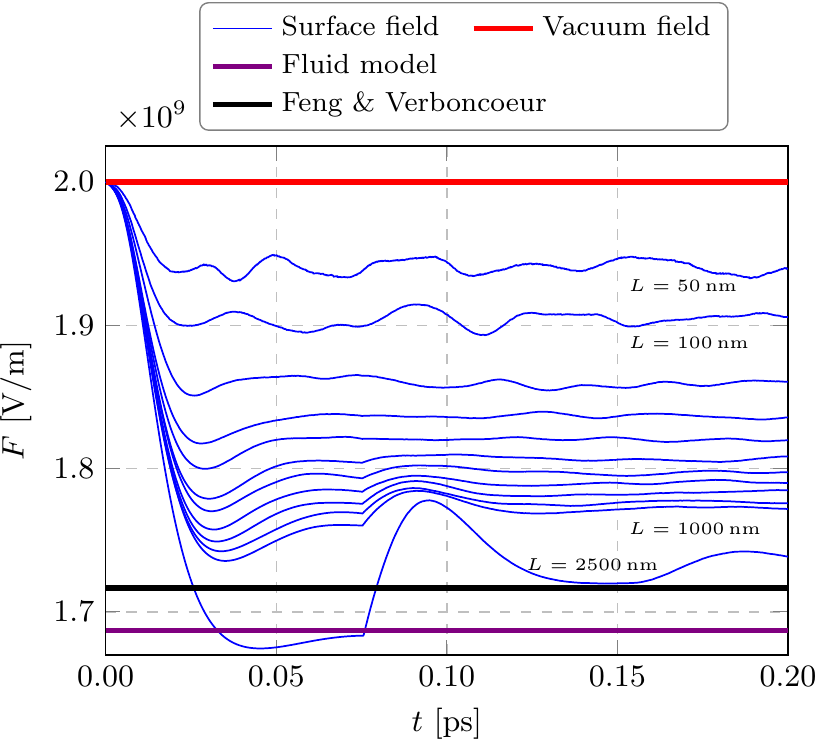}
    \caption{Average surface field plotted for different side lengths, \(L = 50,\, 100,\, 200,\, \ldots,\, 1000,\,2500\,\mathrm{nm}\).
             The magnitude of the steady-state field decreases with increasing \(L\).
             Other parameters were \(\phi = 2.0\,\mathrm{eV}\), \(V = 2\,\mathrm{kV}\) and \(d = 1000\,\mathrm{nm}\). }
    \label{fig:plot-field}
  \end{figure}

  In order to verify the code we use for our simulations, we compared our results to Particle in Cell (PIC) simulations.
  In reference~\onlinecite{Feng2006} Feng and Verboncoeur do 1D PIC simulations to study space-charge limited current with Fowler-Nordheim field emission.
  Using the same parameters as they do, \(\phi = 2.0\,\mathrm{eV}\), \(V = 2\,\mathrm{kV}\) and \(d = 1000\,\mathrm{nm}\),
  we obtain~\autoref{fig:plot-field}, which shows the average surface field on the cathode similar to Fig.~4 in the paper by Feng and Verboncoeur.
  The \textcolor{blue}{blue} lines represent the average surface field for different side lengths \(L\) of the active region on the cathode.
The values used are \(L = 50,\, 100,\, 200,\, \ldots,\,
1000,\,2500\,\mathrm{nm}\). The magnitude of the steady state field
decreases as the side length increases, because the space-charge effects
increase.  For a large side length, comparable to the gap space, the field
resulting from our MD simulations approaches the 1D limit obtained in the steady state
by the PIC method, which we plotted as the \textcolor{black}{black}
line.

The surface field shows time dependent oscillations, both in the
MD and PIC simulations, before it settles down into the steady state.
The oscillations resulting from the PIC method are not shown here (see
Figure 4 of Ref. \cite{Feng2006}), but they are very similar to ours
for \(L=2500\,\mathrm{nm}\) and \(t>0.1\,\mathrm{ps}\).
However, an interesting feature
is present in the MD simulations, but not visible in the PIC results:
it is the sudden increase in the surface field seen at around \(t \approx
0.075\,\mathrm{ps}\), like a kink growing with increasing cathode
size. This time corresponds to the transit time of the electrons in the
system and is the moment when the first electrons are
being absorbed at the anode. The large number of electrons that were
emitted in beginning are now being absorbed which then cause the surface
field to increase abruptly as the absorbed electrons leave the
system. The effect is more pronounced for larger side lengths because
of the higher number of electrons in the system being emitted and absorbed.

The Coulomb oscillations become faster and denser for a small cathode size
because of the larger fluctuations of the emitted charge in combination with
the self consistent space-charge.  Such oscillations may eventually split the
electron beam into bunches if the emission process can be sufficiently fast
as in the photoemission case \cite{PhysRevLett.104.175002}.
In our present approach the diode is initially empty
and a large number of electrons can be emitted in the first time steps.
In this regime, i. e. for \(t<0.01-0.02\,\mathrm{ps}\), the space-charge effects are small. After the
field reaches its lowest magnitude it starts to increase slightly
again because the electron are moving away from the cathode and towards
the anode. The oscillations in the field slowly die away with time.
In our ~\autoref{fig:plot-field} the vacuum field is represented
by the \textcolor{red}{red} line and the \textcolor{violet}{violet}
line shows the equilibrium surface field obtained by the fluid model described in~\autoref{sec:fluid}.

\subsection{Current vs. cathode size}

\begin{figure}[tb]
    \centering
    \includegraphics{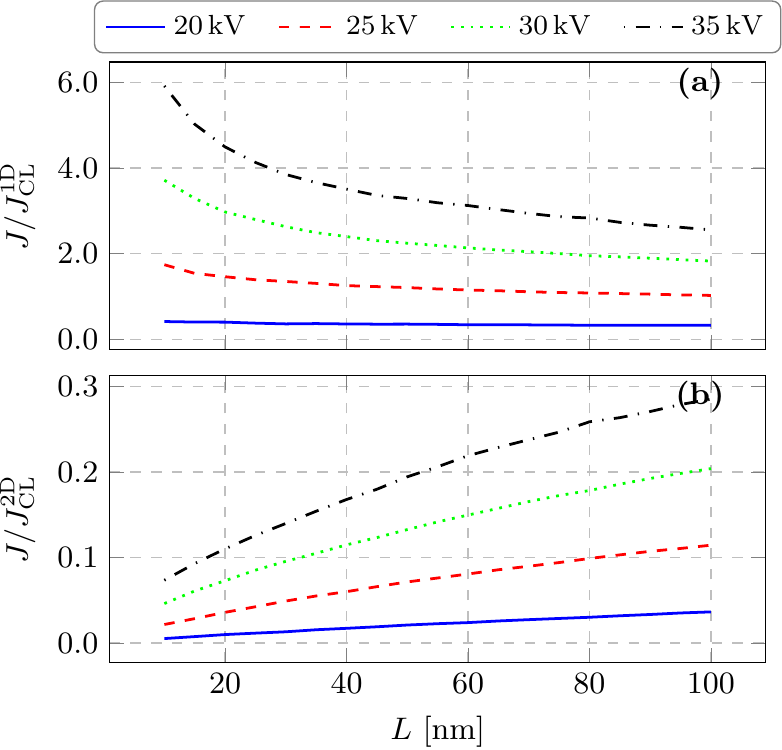}
    \caption{(\textbf{a}) The current density scaled with the 1D Child-Langmuir limit.
             (\textbf{b}) The current density scaled with the 2D Child-Langmuir from Lau~\cite{PhysRevLett.87.278301} as a function of the side length \(L\) for different voltages.
             Other parameters used were \(\Phi = 4.7\,\mathrm{eV}\) and \(d = 2500\,\mathrm{nm}\).}
    \label{fig:JvsLall}
  \end{figure}

  In~\autoref{fig:JvsLall}\hyperref[fig:JvsLall]{(a)} we see the current density scaled with the 1D Child-Langmuir limit
  and in~\autoref{fig:JvsLall}\hyperref[fig:JvsLall]{(b)} scaled with the 2D Child-Langmuir limit from Lau~\cite{PhysRevLett.87.278301}
  as a function of the side length \(L = 10\text{--}100\,\mathrm{nm}\).
  The current density decreases as the size of the active emission
  area on the cathode increases, and approaches an asymptotic value for
  an infinite area.  This can be easily understood since, as is the case with Child-Langmuir emission from a 2D emitter,
  the surface field has its lowest magnitude in the center of the emitter and increases towards
  the edges~\cite{Luginsland2002}. The contribution of the edge
  electrons to the surface field will be less and less as the emitter
  area increases due to the inverse square nature of Coulomb's law.
  The current density will therefore, asymptotically approach some final
  value as the active emission area increases.

\subsection{Beam distribution}

\begin{figure}[tb]
    \centering
    \includegraphics{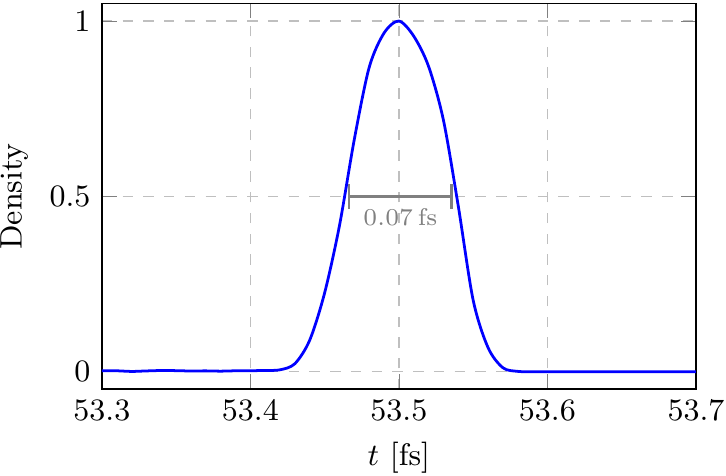}
    \caption{Distribution of the transit time of the electrons through the gap for \(V = 25\,\mathrm{kV}\), \(d = 2500\,\mathrm{nm}\) and \(\phi = 4.7\,\mathrm{eV}\).}
    \label{fig:transit}
  \end{figure}

  \autoref{fig:transit} shows the distribution of the transit time of the electrons through the gap.
  The distribution is approximately a Gaussian with
  a \(\mathrm{FWHM} \approx 0.07\,\mathrm{fs}\), and a peak at \(t =
  53.5\,\mathrm{fs}\).
  From energy conservation the estimated transit time for a single
  electron over the
  gap is \(\Delta t = \sqrt{2 m_e d^2/(V_0
  e)} \approx 53.32\,\mathrm{fs}\), which is not far from the peak
  value of the distribution. It is the Coulomb interaction in the system that slightly
  shifts the peak from the single electron value and
  gives the width of the distribution. The width of the peak is small,
  which indicates that the electrons travel quite fast over the gap and
  do not have a lot of time to interact and spread out.

\begin{figure}[tb]
    \centering
    \includegraphics{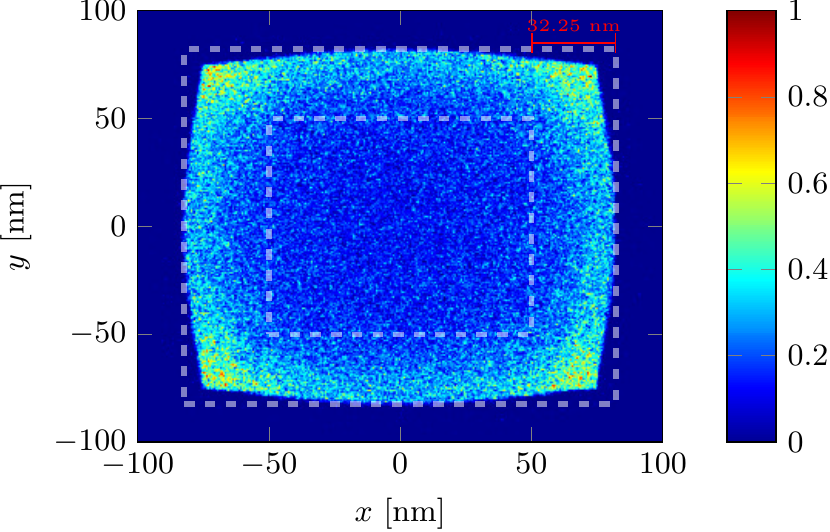}
    \caption{Absorption profile at the anode. Inner box shows the emitter size while the outer box shows the edge of the beam on the absorption plane.
             Parameters used were \(\Phi = 4.7\,\mathrm{eV}\), \(d = 2500\,\mathrm{nm}\) and \(V = 30\,\mathrm{kV}\).}
    \label{fig:abs-30kv}
  \end{figure}

  \autoref{fig:abs-30kv} shows the absorption profile on the anode surface. The inner white square represents the active emission area on the cathode, while the outer white square
  shows the boundaries of the absorbed electrons on the anode. The Coulomb interaction slightly rounds the corners of the beam from the square shape of the emitter.
  The side length of the outer square is about \(2 \times 32.25\,\mathrm{nm}\) larger than the inner square which has \(L = 100\,\mathrm{nm}\). This means that beam spreading
  is small compared with the gap spacing, \({d = 2500\,\mathrm{nm}}\).

\section{Fluid Model\label{sec:fluid}}
  A simpler method that can be used to described the field emission is by
  calculating the electric field in the diode from the charge density.
  The charge density can be estimated by combining the
  continuity equation, \(\rho(z)p/m_e = J\), where \(p\) is the momentum,
  and the conservation of energy \(p^2/(2m_e) = e V_0 z/d\),
  which gives \(\rho(z) = J \sqrt{m_ed/(2eV_0 z)}\).
  Note that we have made use of the vacuum potential to calculate the charge density. This proves to be a
  sufficient approximation for the situation that is being studies where the current density is still
  considerably lower than the 2D Child-Langmuir current density (as can be seen from~\autoref{fig:JvsLall}).  When the
  current density approaches the Child-Langmuir current density an iterative calculation of the potential as
  a function of z can be used instead, at added computational cost.
  This charge density is distributed over the whole diode and therefore it behaves more like a fluid than like
  a collection of single particles. It also assumes the beam does not spread too much laterally, which is a fair
  approximation in our system as can be seen in~\autoref{fig:abs-30kv}.

  Once the charge density is known it is easy to write down the equation
  for the \(z\)-component of the electric field through the center of
  the diode,

  \begin{equation}\label{eq:ez-fluid}
    \!\hat{E}_{\mathrm{sc}}^{\pm}(\hat{z})\! =\! \frac{\hat{J}}{9\pi}\!\!\int\limits_0^1\!\!\! \int\!\!\!\!\!
        \int\limits_{\scalebox{0.5}[1.0]{\( - \)}\frac{L}{2d}}^{\frac{L}{2d}}\!\!\!
          \frac{\hat{z}^\prime \pm \hat{z}}{\sqrt{\hat{z}^\prime} (\hat{x}^{\prime 2} + \hat{y}^{\prime 2} + (\hat{z}^\prime\pm \hat{z})^2)^{\frac{3}{2}}}\,
            \ud \hat{x}^\prime \ud \hat{y}^\prime \ud \hat{z}^\prime ,
  \end{equation}
  where \(\hat{E} = E/(-V_0/d)\), \(\hat{J} = J/J_{\mathrm{CL}}^{1\mathrm{D}}\) and \(\hat{x}\), \(\hat{y}\) and \(\hat{z}\) are scaled using the gap spacing \(d\).
  The plus sign in the integral is used when calculating the image-charge effect and the minus sign for the field in the diode. The total field is then
  \( \hat{E}_z(\hat{z}) = 1 - \hat{E}_{\mathrm{sc}}^+(\hat{z}) - \hat{E}_{\mathrm{sc}}^-(\hat{z})\).

\begin{figure}[tb]
    \centering
    \includegraphics{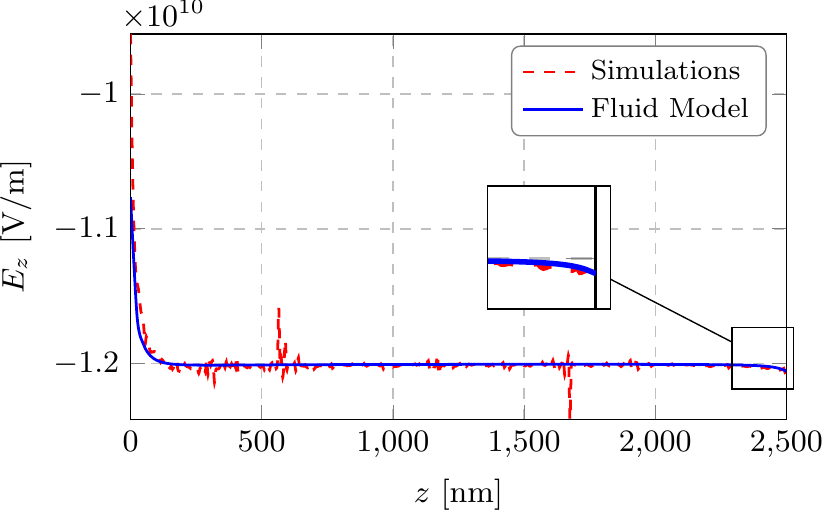}
    \caption{The electric field in the \(z\)-direction through the center of the diode.
             \(V = 30\,\mathrm{kV}\), \(d = 2500\,\mathrm{nm}\) and \(\phi = 4.7\,\mathrm{eV}\).
             Inset shows the electric field near near the anode.}
    \label{fig:Ez-an}
  \end{figure}

  In~\autoref{fig:Ez-an} we see the \(z\)-component of the
  electric field through the center of the diode plotted as a
  function of the \(z\)-coordinate (distance from the cathode). The
  \textcolor{red}{red} dashed curve shows a typical result of an
  MD simulation snapshot at a fixed time step, while
  the \textcolor{blue}{blue} solid curve shows our fluid like model
  calculated using~\autoref{eq:ez-fluid} with numerical integration. The
  value of the current density used, \(J\), is taken from the simulation.
  The fluctuations in the field from the simulations are due to electrons that are close to the center line where the field is being calculated.
%
  The fluid model fits quite well with the simulation results.
  The derivative of the electric field is proportional to the charge density, \(\rho(z)\),
  which is highest near the cathode, where most of the electrons are
  located.  The electrons are injected with zero-velocity and consequently
  spend more time near the cathode before being accelerated over the
  gap. The electrons spread out as their velocity increases, and therefore the charge
  density decreases rapidly and the electric field levels off after the electrons
  have travelled roughly the distance of the side length \(L\) away from the cathode.
  The electric field decreases slightly at the end of the gap, near the anode, due
  to the absorption of the electrons into the anode. There is no charge
  accumulated at the anode, which causes the field to drop slightly. Image-charge
  partners are included in the simulations at the anode, but not in the fluid
  model. It appears that the image-charge contributes little to the field
  near the anode.

  \begin{figure}[tb]
    \centering
    \includegraphics{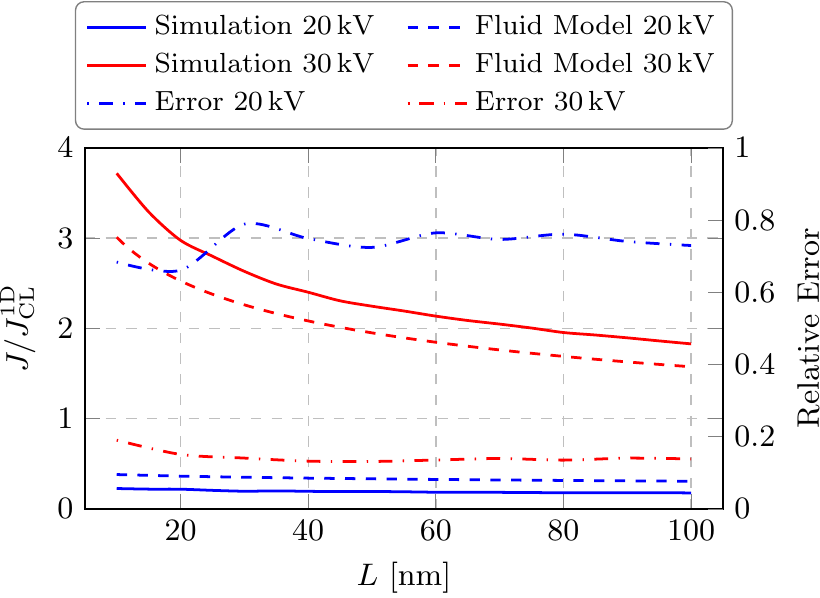}
    \caption{Comparison of the current density between the MD simulation (solid curves) and
             the fluid model (dashed curves) for \(V = 20\,\mathrm{kV}\) (bottom \textcolor{blue}{blue} curves)
             and \(V = 30\,\mathrm{kV}\) (top \textcolor{red}{red} curves).
             Other parameters used were \(d = 2500\,\mathrm{nm}\) and \(\phi = 4.7\,\mathrm{eV}\).
             }
    \label{fig:simvsfluid}
  \end{figure}
  It is also possible to use the fluid model to calculate the current density from the cathode. This is done by iterations, using~\autoref{eq:FN-eq} and~\ref{eq:ez-fluid}.
  The vacuum field is used as an initial value for the field in~\autoref{eq:FN-eq}. The value obtained is then scaled using the Child-Langmuir limit and put
  into~\autoref{eq:ez-fluid} with \(z = 0\), which gives a new value for the surface field in center of the cathode. This procedure is repeated until the
  current density has converged.

  In~\autoref{fig:simvsfluid} we see a comparison between the fluid model and the simulation results for calculating
  the current density emitted. The \textcolor{blue}{blue} solid curve shows the simulations results, while the \textcolor{red}{red} dashed the fluid model calculate
  using method just described.
  It is expected that the fluid model would
  give slightly lower results. The model calculates the surface field in the center of the diode and as was explained earlier the field is lower there than at the edges.
  The fluid model therefore underestimates the current density emitted from the edges.

\begin{figure}[tb]
    \centering
    \includegraphics{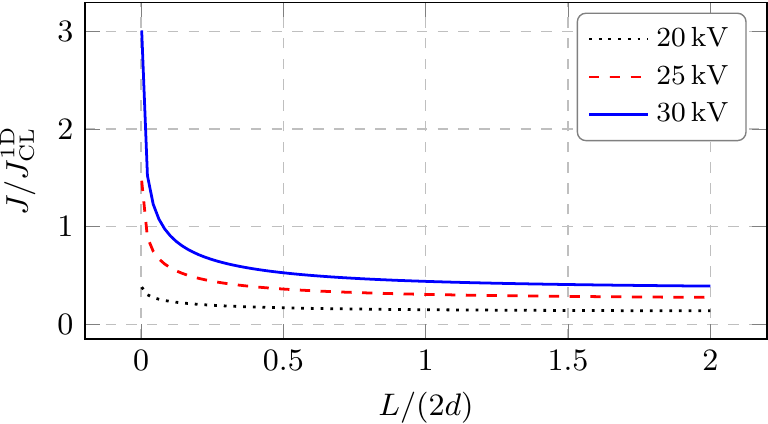}
    \caption{Results of the fluid model for different values of the voltage as a function of the side length.
             \(L = 10\text{--}10.000\,\mathrm{nm}\) with \(d = 2500\,\mathrm{nm}\) and \(\phi = 4.7\mathrm{eV}\).}
    \label{fig:fluid-model}
  \end{figure}

  In~\autoref{fig:fluid-model} we see the results of the fluid model for different values of the voltage as a function of
  \({L = 10\text{--}10.000\,\mathrm{nm}}\) with \(d = 2500\,\mathrm{nm}\) and \(\phi = 4.7\mathrm{eV}\).
  The results are qualitatively the same as in~\autoref{fig:JvsLall}\hyperref[fig:JvsLall]{(b)}.
  The fluid model allows us to estimate the current density for diodes with a larger emitting area and see the asymptotic behaviour better with calculations that
  can be done in a few hours, whereas the MD simulations would require weeks of computational time for the same parameters. The shape of the curves seen is very similar
  to the current profile seen in space-charge limited flow from circular cathodes~\cite{Kelley13243474}.

\section{Summary and conclusions\label{sec:summary}}

We performed molecular dynamics (MD) simulations of electrons in a diode
where the electrons are extracted from the cathode by the field emission
mechanism.  The emission is essentially governed by the quantum mechanical
tunnelling through the potential barrier associated with the surface of the
metallic cathode, which leads to the well known Fowler-Nordheim law for
an infinite planar diode.

Our results include space-charge effects and are in good agreement
with the results obtained by other authors~\cite{Feng2006, Rokhlenko3272690}.
In particular they are in good agreement with the PIC simulations.
MD simulations are more computationally expensive than PIC simulations,
but are more accurate.
In the MD simulation every particle is tracked precisely and the force on it is
calculated exactly, using the Coulomb interaction from all other particles
in the system. Whereas in PIC simulation the system is divided up into
a grid and the field in each section of the grid is the mean field of
the particles in that section. The only methodological approximation in
the MD simulations is the finite time step, which must be chosen much
smaller than the characteristic time of the beam dynamics.

Time dependent oscillations of the electric field are obtained during
the transient period after the diode is switched on. A kink of cathode
field is obtained as a response to the charge fluctuation produced when
the first electrons are absorbed at the anode.

The MD method is best suited when the number of the electrons in the diode is
not too large, typically below few thousands. For larger numbers it
becomes computationally prohibited, but it those cases the mean field results
like those of PIC calculations, are usually accurate.

We derive a relatively simple fluid model of the electron beam,
which incorporates the essential electrostatics, where the
electric field is derived self consistently with an estimated continuous
charge distribution, and with the image-charge induced at the cathode and at the anode.
The results of the fluid model are good agreement with those of the MD simulations.
We can use this fluid model to estimate the effects of finite emitter area over a wide range of parameters.
We observe that the current density for field emission from a finite emitter with space-charge effects included,
is qualitatively similar to Child-Langmuir emission from a finite emitter area,
in that the current density increases with diminishing emitter area.

\begin{acknowledgments}
  This work was financially supported by the Icelandic Research Fund grant number 120009021.
  The simulations were performed on resources provided by the Nordic High Performance Computing (NHPC).
\end{acknowledgments}

\bibliography{bibliography}

\begin{thebibliography}{21}%
\makeatletter
\providecommand \@ifxundefined [1]{%
 \@ifx{#1\undefined}
}%
\providecommand \@ifnum [1]{%
 \ifnum #1\expandafter \@firstoftwo
 \else \expandafter \@secondoftwo
 \fi
}%
\providecommand \@ifx [1]{%
 \ifx #1\expandafter \@firstoftwo
 \else \expandafter \@secondoftwo
 \fi
}%
\providecommand \natexlab [1]{#1}%
\providecommand \enquote  [1]{``#1''}%
\providecommand \bibnamefont  [1]{#1}%
\providecommand \bibfnamefont [1]{#1}%
\providecommand \citenamefont [1]{#1}%
\providecommand \href@noop [0]{\@secondoftwo}%
\providecommand \href [0]{\begingroup \@sanitize@url \@href}%
\providecommand \@href[1]{\@@startlink{#1}\@@href}%
\providecommand \@@href[1]{\endgroup#1\@@endlink}%
\providecommand \@sanitize@url [0]{\catcode `\\12\catcode `\$12\catcode
  `\&12\catcode `\#12\catcode `\^12\catcode `\_12\catcode `\%12\relax}%
\providecommand \@@startlink[1]{}%
\providecommand \@@endlink[0]{}%
\providecommand \url  [0]{\begingroup\@sanitize@url \@url }%
\providecommand \@url [1]{\endgroup\@href {#1}{\urlprefix }}%
\providecommand \urlprefix  [0]{URL }%
\providecommand \Eprint [0]{\href }%
\providecommand \doibase [0]{http://dx.doi.org/}%
\providecommand \selectlanguage [0]{\@gobble}%
\providecommand \bibinfo  [0]{\@secondoftwo}%
\providecommand \bibfield  [0]{\@secondoftwo}%
\providecommand \translation [1]{[#1]}%
\providecommand \BibitemOpen [0]{}%
\providecommand \bibitemStop [0]{}%
\providecommand \bibitemNoStop [0]{.\EOS\space}%
\providecommand \EOS [0]{\spacefactor3000\relax}%
\providecommand \BibitemShut  [1]{\csname bibitem#1\endcsname}%
\let\auto@bib@innerbib\@empty
\bibitem [{\citenamefont {Lau}, \citenamefont {Liu},\ and\ \citenamefont
  {Parker}(1994)}]{Lau1870603}%
  \BibitemOpen
  \bibfield  {author} {\bibinfo {author} {\bibfnamefont {Y.~Y.}\ \bibnamefont
  {Lau}}, \bibinfo {author} {\bibfnamefont {Y.}~\bibnamefont {Liu}}, \ and\
  \bibinfo {author} {\bibfnamefont {R.~K.}\ \bibnamefont {Parker}},\ }\href
  {\doibase 10.1063/1.870603} {\bibfield  {journal} {\bibinfo  {journal}
  {Physics of Plasmas (1994-present)}\ }\textbf {\bibinfo {volume} {1}},\
  \bibinfo {pages} {2082} (\bibinfo {year} {1994})}\BibitemShut {NoStop}%
\bibitem [{\citenamefont {Feng}\ and\ \citenamefont
  {Verboncoeur}(2006)}]{Feng2006}%
  \BibitemOpen
  \bibfield  {author} {\bibinfo {author} {\bibfnamefont {Y.}~\bibnamefont
  {Feng}}\ and\ \bibinfo {author} {\bibfnamefont {J.~P.}\ \bibnamefont
  {Verboncoeur}},\ }\href {\doibase 10.1063/1.2226977} {\bibfield  {journal}
  {\bibinfo  {journal} {Physics of Plasmas (1994-present)}\ }\textbf {\bibinfo
  {volume} {13}},\ \bibinfo {eid} {073105} (\bibinfo {year}
  {2006})}\BibitemShut {NoStop}%
\bibitem [{\citenamefont {Fowler}\ and\ \citenamefont
  {Nordheim}(1928)}]{Fowler01051928}%
  \BibitemOpen
  \bibfield  {author} {\bibinfo {author} {\bibfnamefont {R.~H.}\ \bibnamefont
  {Fowler}}\ and\ \bibinfo {author} {\bibfnamefont {L.}~\bibnamefont
  {Nordheim}},\ }\href {\doibase 10.1098/rspa.1928.0091} {\bibfield  {journal}
  {\bibinfo  {journal} {Proceedings of the Royal Society of London. Series A}\
  }\textbf {\bibinfo {volume} {119}},\ \bibinfo {pages} {173} (\bibinfo {year}
  {1928})}\BibitemShut {NoStop}%
\bibitem [{\citenamefont {Forbes}\ and\ \citenamefont
  {Deane}(2007)}]{Forbes08112007}%
  \BibitemOpen
  \bibfield  {author} {\bibinfo {author} {\bibfnamefont {R.~G.}\ \bibnamefont
  {Forbes}}\ and\ \bibinfo {author} {\bibfnamefont {J.~H.}\ \bibnamefont
  {Deane}},\ }\href {\doibase 10.1098/rspa.2007.0030} {\bibfield  {journal}
  {\bibinfo  {journal} {Proceedings of the Royal Society A: Mathematical,
  Physical and Engineering Science}\ }\textbf {\bibinfo {volume} {463}},\
  \bibinfo {pages} {2907} (\bibinfo {year} {2007})}\BibitemShut {NoStop}%
\bibitem [{\citenamefont {Jensen}(2003)}]{Jensen1528}%
  \BibitemOpen
  \bibfield  {author} {\bibinfo {author} {\bibfnamefont {K.~L.}\ \bibnamefont
  {Jensen}},\ }\href {\doibase 10.1116/1.1573664} {\bibfield  {journal}
  {\bibinfo  {journal} {Journal of Vacuum Science \& Technology B}\ }\textbf
  {\bibinfo {volume} {21}},\ \bibinfo {pages} {1528} (\bibinfo {year}
  {2003})}\BibitemShut {NoStop}%
\bibitem [{\citenamefont {Eichmeier}\ and\ \citenamefont
  {Thumm}(2008)}]{eichmeier2008vacuum}%
  \BibitemOpen
  \bibinfo {editor} {\bibfnamefont {J.~A.}\ \bibnamefont {Eichmeier}}\ and\
  \bibinfo {editor} {\bibfnamefont {M.}~\bibnamefont {Thumm}},\ eds.,\
  \href@noop {} {\emph {\bibinfo {title} {Vacuum electronics: Components and
  Devices}}}\ (\bibinfo  {publisher} {Springer},\ \bibinfo {year}
  {2008})\BibitemShut {NoStop}%
\bibitem [{\citenamefont {Zhu}(2004)}]{zhu2004vacuum}%
  \BibitemOpen
  \bibinfo {editor} {\bibfnamefont {W.}~\bibnamefont {Zhu}},\ ed.,\ \href@noop
  {} {\emph {\bibinfo {title} {Vacuum Microelectronics}}}\ (\bibinfo
  {publisher} {John Wiley \& Sons},\ \bibinfo {year} {2004})\BibitemShut
  {NoStop}%
\bibitem [{\citenamefont {Neuber}\ \emph {et~al.}(2001)\citenamefont {Neuber},
  \citenamefont {Laurent}, \citenamefont {Lau},\ and\ \citenamefont
  {Krompholz}}]{neuber2001windows}%
  \BibitemOpen
  \bibfield  {author} {\bibinfo {author} {\bibfnamefont {A.}~\bibnamefont
  {Neuber}}, \bibinfo {author} {\bibfnamefont {L.}~\bibnamefont {Laurent}},
  \bibinfo {author} {\bibfnamefont {Y.}~\bibnamefont {Lau}}, \ and\ \bibinfo
  {author} {\bibfnamefont {H.}~\bibnamefont {Krompholz}},\ }\href@noop {}
  {\emph {\bibinfo {title} {Windows and RF Breakdown}}},\ edited by\ \bibinfo
  {editor} {\bibfnamefont {R.}~\bibnamefont {Barker}}\ and\ \bibinfo {editor}
  {\bibfnamefont {E.}~\bibnamefont {Schlamigolu}}\ (\bibinfo  {publisher}
  {Wiley-IEEE Press},\ \bibinfo {year} {2001})\ pp.\ \bibinfo {pages}
  {325--375},\ \bibinfo {note} {chapter 10 in High-Power Microwave Sources and
  Technologies}\BibitemShut {NoStop}%
\bibitem [{\citenamefont {Barbour}\ \emph {et~al.}(1953)\citenamefont
  {Barbour}, \citenamefont {Dolan}, \citenamefont {Trolan}, \citenamefont
  {Martin},\ and\ \citenamefont {Dyke}}]{PhysRev.92.45}%
  \BibitemOpen
  \bibfield  {author} {\bibinfo {author} {\bibfnamefont {J.~P.}\ \bibnamefont
  {Barbour}}, \bibinfo {author} {\bibfnamefont {W.~W.}\ \bibnamefont {Dolan}},
  \bibinfo {author} {\bibfnamefont {J.~K.}\ \bibnamefont {Trolan}}, \bibinfo
  {author} {\bibfnamefont {E.~E.}\ \bibnamefont {Martin}}, \ and\ \bibinfo
  {author} {\bibfnamefont {W.~P.}\ \bibnamefont {Dyke}},\ }\href {\doibase
  10.1103/PhysRev.92.45} {\bibfield  {journal} {\bibinfo  {journal} {Phys.
  Rev.}\ }\textbf {\bibinfo {volume} {92}},\ \bibinfo {pages} {45} (\bibinfo
  {year} {1953})}\BibitemShut {NoStop}%
\bibitem [{\citenamefont {Rokhlenko}, \citenamefont {Jensen},\ and\
  \citenamefont {Lebowitz}(2010)}]{Rokhlenko3272690}%
  \BibitemOpen
  \bibfield  {author} {\bibinfo {author} {\bibfnamefont {A.}~\bibnamefont
  {Rokhlenko}}, \bibinfo {author} {\bibfnamefont {K.~L.}\ \bibnamefont
  {Jensen}}, \ and\ \bibinfo {author} {\bibfnamefont {J.~L.}\ \bibnamefont
  {Lebowitz}},\ }\href {\doibase 10.1063/1.3272690} {\bibfield  {journal}
  {\bibinfo  {journal} {Journal of Applied Physics}\ }\textbf {\bibinfo
  {volume} {107}},\ \bibinfo {eid} {014904} (\bibinfo {year}
  {2010})}\BibitemShut {NoStop}%
\bibitem [{\citenamefont {Jensen}(2010)}]{Jensen14905}%
  \BibitemOpen
  \bibfield  {author} {\bibinfo {author} {\bibfnamefont {K.~L.}\ \bibnamefont
  {Jensen}},\ }\href {\doibase 10.1063/1.3272688} {\bibfield  {journal}
  {\bibinfo  {journal} {Journal of Applied Physics}\ }\textbf {\bibinfo
  {volume} {107}},\ \bibinfo {eid} {014905} (\bibinfo {year}
  {2010})}\BibitemShut {NoStop}%
\bibitem [{\citenamefont {Jensen}\ \emph {et~al.}(2012)\citenamefont {Jensen},
  \citenamefont {Lebowitz}, \citenamefont {Lau},\ and\ \citenamefont
  {Luginsland}}]{Jensen3692577}%
  \BibitemOpen
  \bibfield  {author} {\bibinfo {author} {\bibfnamefont {K.~L.}\ \bibnamefont
  {Jensen}}, \bibinfo {author} {\bibfnamefont {J.}~\bibnamefont {Lebowitz}},
  \bibinfo {author} {\bibfnamefont {Y.~Y.}\ \bibnamefont {Lau}}, \ and\
  \bibinfo {author} {\bibfnamefont {J.}~\bibnamefont {Luginsland}},\ }\href
  {\doibase 10.1063/1.3692577} {\bibfield  {journal} {\bibinfo  {journal}
  {Journal of Applied Physics}\ }\textbf {\bibinfo {volume} {111}},\ \bibinfo
  {eid} {054917} (\bibinfo {year} {2012})}\BibitemShut {NoStop}%
\bibitem [{\citenamefont {Pedersen}, \citenamefont {Manolescu},\ and\
  \citenamefont {Valfells}(2010)}]{PhysRevLett.104.175002}%
  \BibitemOpen
  \bibfield  {author} {\bibinfo {author} {\bibfnamefont {A.}~\bibnamefont
  {Pedersen}}, \bibinfo {author} {\bibfnamefont {A.}~\bibnamefont {Manolescu}},
  \ and\ \bibinfo {author} {\bibfnamefont {A.}~\bibnamefont {Valfells}},\
  }\href {\doibase 10.1103/PhysRevLett.104.175002} {\bibfield  {journal}
  {\bibinfo  {journal} {Phys. Rev. Lett.}\ }\textbf {\bibinfo {volume} {104}},\
  \bibinfo {pages} {175002} (\bibinfo {year} {2010})}\BibitemShut {NoStop}%
\bibitem [{\citenamefont {Jonsson}\ \emph {et~al.}(2013)\citenamefont
  {Jonsson}, \citenamefont {Ilkov}, \citenamefont {Manolescu}, \citenamefont
  {Pedersen},\ and\ \citenamefont {Valfells}}]{Jonsson4793451}%
  \BibitemOpen
  \bibfield  {author} {\bibinfo {author} {\bibfnamefont {P.}~\bibnamefont
  {Jonsson}}, \bibinfo {author} {\bibfnamefont {M.}~\bibnamefont {Ilkov}},
  \bibinfo {author} {\bibfnamefont {A.}~\bibnamefont {Manolescu}}, \bibinfo
  {author} {\bibfnamefont {A.}~\bibnamefont {Pedersen}}, \ and\ \bibinfo
  {author} {\bibfnamefont {A.}~\bibnamefont {Valfells}},\ }\href {\doibase
  10.1063/1.4793451} {\bibfield  {journal} {\bibinfo  {journal} {Physics of
  Plasmas (1994-present)}\ }\textbf {\bibinfo {volume} {20}},\ \bibinfo {eid}
  {023107} (\bibinfo {year} {2013})}\BibitemShut {NoStop}%
\bibitem [{\citenamefont {Ilkov}\ \emph {et~al.}(2015)\citenamefont {Ilkov},
  \citenamefont {Torfason}, \citenamefont {Manolescu},\ and\ \citenamefont
  {Valfells}}]{6979259}%
  \BibitemOpen
  \bibfield  {author} {\bibinfo {author} {\bibfnamefont {M.}~\bibnamefont
  {Ilkov}}, \bibinfo {author} {\bibfnamefont {K.}~\bibnamefont {Torfason}},
  \bibinfo {author} {\bibfnamefont {A.}~\bibnamefont {Manolescu}}, \ and\
  \bibinfo {author} {\bibfnamefont {A.}~\bibnamefont {Valfells}},\ }\href
  {\doibase 10.1109/TED.2014.2370680} {\bibfield  {journal} {\bibinfo
  {journal} {Electron Devices, IEEE Transactions on}\ }\textbf {\bibinfo
  {volume} {62}},\ \bibinfo {pages} {200} (\bibinfo {year} {2015})}\BibitemShut
  {NoStop}%
\bibitem [{\citenamefont {Hastings}(1970)}]{Hastings}%
  \BibitemOpen
  \bibfield  {author} {\bibinfo {author} {\bibfnamefont {W.~K.}\ \bibnamefont
  {Hastings}},\ }\href {http://www.jstor.org/stable/2334940} {\bibfield
  {journal} {\bibinfo  {journal} {Biometrika}\ }\textbf {\bibinfo {volume}
  {57}},\ \bibinfo {pages} {pp. 97} (\bibinfo {year} {1970})}\BibitemShut
  {NoStop}%
\bibitem [{\citenamefont {Child}(1911)}]{PhysRevSeriesI.32.492}%
  \BibitemOpen
  \bibfield  {author} {\bibinfo {author} {\bibfnamefont {C.~D.}\ \bibnamefont
  {Child}},\ }\href {\doibase 10.1103/PhysRevSeriesI.32.492} {\bibfield
  {journal} {\bibinfo  {journal} {Phys. Rev. (Series I)}\ }\textbf {\bibinfo
  {volume} {32}},\ \bibinfo {pages} {492} (\bibinfo {year} {1911})}\BibitemShut
  {NoStop}%
\bibitem [{\citenamefont {Langmuir}(1913)}]{PhysRev.2.450}%
  \BibitemOpen
  \bibfield  {author} {\bibinfo {author} {\bibfnamefont {I.}~\bibnamefont
  {Langmuir}},\ }\href {\doibase 10.1103/PhysRev.2.450} {\bibfield  {journal}
  {\bibinfo  {journal} {Phys. Rev.}\ }\textbf {\bibinfo {volume} {2}},\
  \bibinfo {pages} {450} (\bibinfo {year} {1913})}\BibitemShut {NoStop}%
\bibitem [{\citenamefont {Lau}(2001)}]{PhysRevLett.87.278301}%
  \BibitemOpen
  \bibfield  {author} {\bibinfo {author} {\bibfnamefont {Y.~Y.}\ \bibnamefont
  {Lau}},\ }\href {\doibase 10.1103/PhysRevLett.87.278301} {\bibfield
  {journal} {\bibinfo  {journal} {Phys. Rev. Lett.}\ }\textbf {\bibinfo
  {volume} {87}},\ \bibinfo {pages} {278301} (\bibinfo {year}
  {2001})}\BibitemShut {NoStop}%
\bibitem [{\citenamefont {Luginsland}\ \emph {et~al.}(2002)\citenamefont
  {Luginsland}, \citenamefont {Lau}, \citenamefont {Umstattd},\ and\
  \citenamefont {Watrous}}]{Luginsland2002}%
  \BibitemOpen
  \bibfield  {author} {\bibinfo {author} {\bibfnamefont {J.~W.}\ \bibnamefont
  {Luginsland}}, \bibinfo {author} {\bibfnamefont {Y.~Y.}\ \bibnamefont {Lau}},
  \bibinfo {author} {\bibfnamefont {R.~J.}\ \bibnamefont {Umstattd}}, \ and\
  \bibinfo {author} {\bibfnamefont {J.~J.}\ \bibnamefont {Watrous}},\ }\href
  {\doibase 10.1063/1.1459453} {\bibfield  {journal} {\bibinfo  {journal}
  {Physics of Plasmas (1994-present)}\ }\textbf {\bibinfo {volume} {9}},\
  \bibinfo {pages} {2371} (\bibinfo {year} {2002})}\BibitemShut {NoStop}%
\bibitem [{\citenamefont {Ragan-Kelley}, \citenamefont {Verboncoeur},\ and\
  \citenamefont {Feng}(2009)}]{Kelley13243474}%
  \BibitemOpen
  \bibfield  {author} {\bibinfo {author} {\bibfnamefont {B.}~\bibnamefont
  {Ragan-Kelley}}, \bibinfo {author} {\bibfnamefont {J.}~\bibnamefont
  {Verboncoeur}}, \ and\ \bibinfo {author} {\bibfnamefont {Y.}~\bibnamefont
  {Feng}},\ }\href {\doibase 10.1063/1.3243474} {\bibfield  {journal} {\bibinfo
   {journal} {Physics of Plasmas (1994-present)}\ }\textbf {\bibinfo {volume}
  {16}},\ \bibinfo {eid} {103102} (\bibinfo {year} {2009})}\BibitemShut
  {NoStop}%
\end{thebibliography}%

\end{document}